\documentclass{webofc}

\usepackage[varg]{txfonts}   
\usepackage{hyperref}
\usepackage{url}
\hypersetup{colorlinks=true,citecolor=blue,urlcolor=blue,linkcolor=blue}
\begin{document}
\title{Imaging the Jet-Induced Medium Response with Energy Correlators}
%
%

\author{\firstname{Hannah} \lastname{Bossi}\inst{1}
\and
\firstname{Arjun} \lastname{Kudinoor}\inst{2}\fnsep\thanks{Speaker at QM25; \email{kudinoor@mit.edu}}
\and
\firstname{Ian} \lastname{Moult}\inst{3}
\and
\firstname{Daniel} \lastname{Pablos}\inst{4,5,6}
\and
\firstname{Ananya} \lastname{Rai}\inst{3}
\and
\firstname{Krishna} \lastname{Rajagopal}\inst{2}
}

\institute{Massachusetts Institute of Technology, Cambridge, MA, 02139
\and
Center for Theoretical Physics -- a Leinweber Institute, MIT, Cambridge, MA 02139
\and
Department of Physics, Yale University, New Haven, CT 06511
\and
IGFAE, Universidade de Santiago de Compostela, E-15782 Spain
\and
Departamento de F\'isica, Universidad de Oviedo, 33007 Oviedo, Spain
\and
ICTEA, Calle de la Independencia 13, 33004 Oviedo, Spain
}

\abstract{
Quark-gluon plasma (QGP), when viewed at length scales of order the inverse of its temperature, behaves as a strongly-coupled liquid. However, when it is probed with sufficiently high momentum transfer, asymptotic freedom mandates the presence of quark- and gluon-like quasi-particles. High energy partons within jets can trigger these high-momentum exchanges, making jets valuable probes for revealing the presence of such quasi-particles. Such elastic scatterings are implemented in the Hybrid Model, where a jet parton that scatters is deflected, kicking a medium parton, which recoils. Before and after a scattering, as one and then both partons propagate through the medium, they lose energy and momentum, exciting wakes in the QGP droplet. We use two-point and three-point energy-energy correlators (EECs) to reveal the relevant angular regions at which (modified) parton showers and wakes in the QGP each dominate, offering a new way with which to visualize and constrain the corresponding dynamics. We compare our calculations to recent CMS and ALICE measurements of two-point EECs of charged-particle tracks in jets produced in PbPb collisions. We show that our calculations are closest to the experimental measurements when elastic scattering is included and when the elastically scattered recoil-partons produce their own wakes. We also propose a new variant of the measurement that is especially sensitive to jet wakes.
}
\maketitle
\section{The Hybrid Strong/Weak Coupling Model}
\label{hybrid-model}
The hybrid strong/weak coupling model, or simply the Hybrid Model, is a theoretical framework for jet quenching in heavy ion collisions.
Jet showers emerge from parton splittings that are determined using the high-$Q^2$ perturbative DGLAP equations, implemented using PYTHIA8. Meanwhile, soft momentum exchanges between  partons in the jet shower and the droplet of quark-gluon plasma (QGP) through which they propagate cause these partons to lose energy. In the Hybrid Model, each parton in a jet shower loses energy to the plasma as determined by a holographic energy loss formula,
detailed in Refs.~\cite{Casalderrey-Solana:2014bpa,Bossi:2024qho,Kudinoor:2025ilx}. The momentum and energy lost by a parton is deposited into the plasma, exciting a hydrodynamic wake within the droplet of QGP. In the Hybrid Model, jet-induced wakes are implemented by generating soft hadrons according to a spectrum determined by employing the Cooper--Frye prescription to the jet-induced perturbation of the stress-energy tensor of the liquid QGP~\cite{Casalderrey-Solana:2016jvj}.
The shape of the wake-spectrum and the assumptions in its calculation are discussed in Refs.~\cite{Casalderrey-Solana:2016jvj,Bossi:2024qho,Kudinoor:2025ilx}.

$2 \rightarrow 2$ elastic (Molière) scatterings are another channel for high-energy jet-partons to interact with the medium.
When QGP is probed at sufficiently high momentum exchange, asymptotic freedom mandates the presence of quark- and gluon-like quasi-particles.
High energy jet-partons can trigger these high-momentum exchanges, making jets valuable probes for revealing the presence of such quasi-particles. Within the Hybrid Model, a jet parton that scatters is deflected, kicking a medium parton, which recoils. As both of these partons propagate further through the medium they lose energy and momentum to the medium, exciting wakes in the medium, as described in Ref.~\cite{Hulcher:2022kmn}. That is, elastic scatterings modify both the parton shower and the wake that the shower excites in the droplet of QGP.

\section{Two-point Energy Correlators}
\begin{figure*}
\centering
\includegraphics[width=13cm,clip]{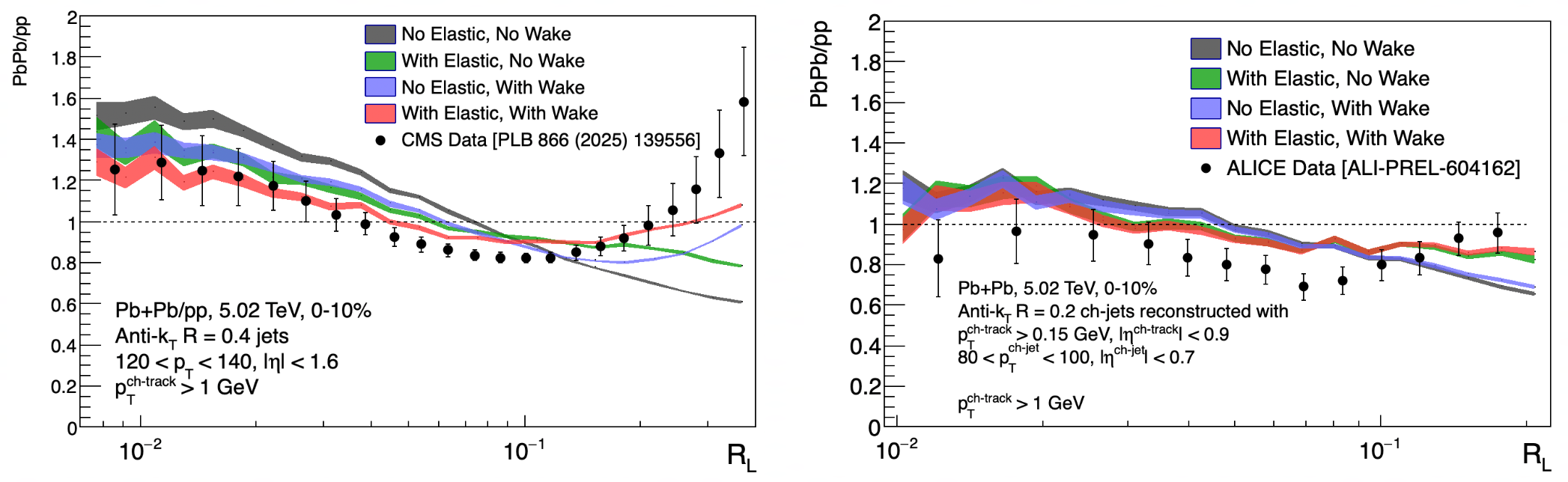}
\caption{PbPb/pp ratio of EECs calculated using charged-particle tracks with $p_{\rm T}^{\rm ch} > 1$ GeV within a sample of inclusive jets with $120 < p_{\rm T} < 140$ GeV $R = 0.4$ (left) and $80 < p_{\rm T}^{\rm ch} < 100$ GeV $R = 0.2$ charged-jets (right). We plot colored Hybrid Model curves, with/without elastic scattering, and with/without hadrons from the wakes of jets. CMS (left) and ALICE (right) experimental measurements from Refs.~\cite{CMS:2025ydi, ALI-PREL-604162} are depicted using point markers. The vertical bars on the experimental
data points indicate the combined statistical and systematic uncertainties, added in quadrature.}
\label{fig:eec}       
\end{figure*}

Recent years have seen substantial advances in our understanding of jet substructure through the use of energy 
correlators (EECs)~\cite{Moult:2025nhu}.
These observables are now being applied in heavy-ion collisions to provide new insight into the interplay between perturbative parton showers and medium-induced phenomena.

\begin{figure}[h]
\centering
\includegraphics[width=8cm,clip]{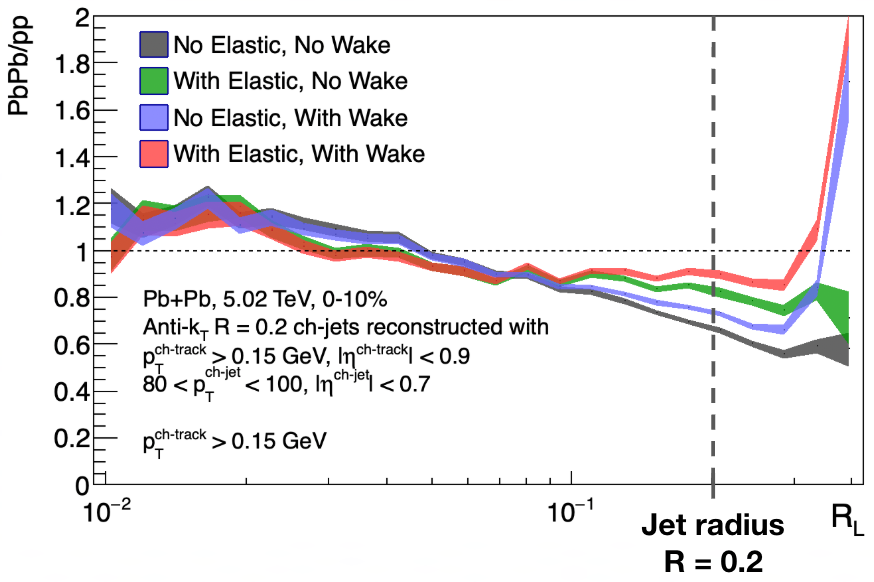}
\caption{PbPb/pp ratio of charged-particle EECs calculated using a sample of $80 < p_{\rm T}^{\rm ch} < 100$ GeV $R = 0.2$ charged-jets and a particle track-$p_{\rm T}$ cut of $150$ MeV. Hybrid Model curves are shown with/without elastic scattering and hadrons from the wakes of jets. The horizontal $R_{\rm L}$-axis extends up to twice the jet radius. Our results motivate extending the EEC measurement in this way.}
\label{fig:full-eec}
\end{figure}

Fig.~\ref{fig:eec} shows the PbPb/pp ratio of EECs, calculated using charged particle tracks with $p_{\rm T}^{\rm ch} > 1$ GeV, for two representative jet selections: inclusive $R=0.4$ jets with $120<p_{\rm T}<140$~GeV (left) and charged-particle $R=0.2$ jets with $80<p_{\rm T}^{\rm ch-jet}<100$~GeV (right). Our model calculations are compared to experimental measurements by CMS~\cite{CMS:2025ydi} (left) and ALICE~\cite{ALI-PREL-604162} (right). Elastic scatterings off quasi-particles in the medium angularly deflect energetic jet-partons. This nudges some of the constituent momenta of jets away from their hard, central cores, enhancing the PbPb/pp EEC ratio in the region $0.1\lesssim R_{\rm L}\lesssim 0.2$, as seen in the right panel of Fig.~\ref{fig:eec}. Crucially, both the deflected jet parton and the recoiling medium parton subsequently lose energy and each excites its own broad wake.

Hybrid Model calculations that include \emph{both} wakes and elastic scatterings track the observed rise in the PbPb/pp ratio toward larger $R_L$ most closely. However, excluding either the wake or elastic scatterings degrades agreement with the experimental data.
Turning off elastic scatterings suppresses medium-induced angular deflections of otherwise collinear partons in the hard cores of jets; turning off the wake removes the broad distribution of soft hadrons responsible for the rise in the PbPb/pp ratio towards large-$R_L$, in particular for $R=0.4$. The importance of the wakes of scattered partons for $R_L>0.2$ in the left panel of Fig.~\ref{fig:eec} tells us that for the (many) scattered jet partons that subsequently lose most or all of their energy, the memory of their deflection is encoded in modifications to the wakes of their jets.

We propose a new and incisive variant of ALICE's measurement of $R=0.2$ charged-jet EECs (from Fig.~\ref{fig:eec}, right)  in Fig.~\ref{fig:full-eec}, where EECs are calculated using a lower track-$p_{\rm T}$ threshold of $p_{\rm T}^{\rm ch}>150$~MeV and the $R_L$-axis is extended beyond the jet radius to $2R$. It is infeasible to obtain pairwise angular separations larger than the jet radius between two partons if one of them resides in the central core of the jet. So, the region $R_L\gtrsim R$ excludes correlations between the hard degrees of freedom inside the jet-core and favors correlations among the soft, peripheral degrees of freedom in the medium-response.
The marked enhancement of the PbPb/pp ratio in Fig.~\ref{fig:full-eec} in the region $R_L > R$ in our Hybrid Model calculations is therefore a direct signal of the wake, motivating extending the EEC measurement to $R_L>R$.

\section{Three-point Energy Correlators}
Studying the EEC allows for the identification of the angular scale of jet-medium interactions. Higher-point correlators hold the promise of mapping out the dynamics themselves. In this section, we focus on the three-point EEEC, constructed from triplets of particles within jets.

\begin{figure*}
\centering
\includegraphics[width=10cm,clip]{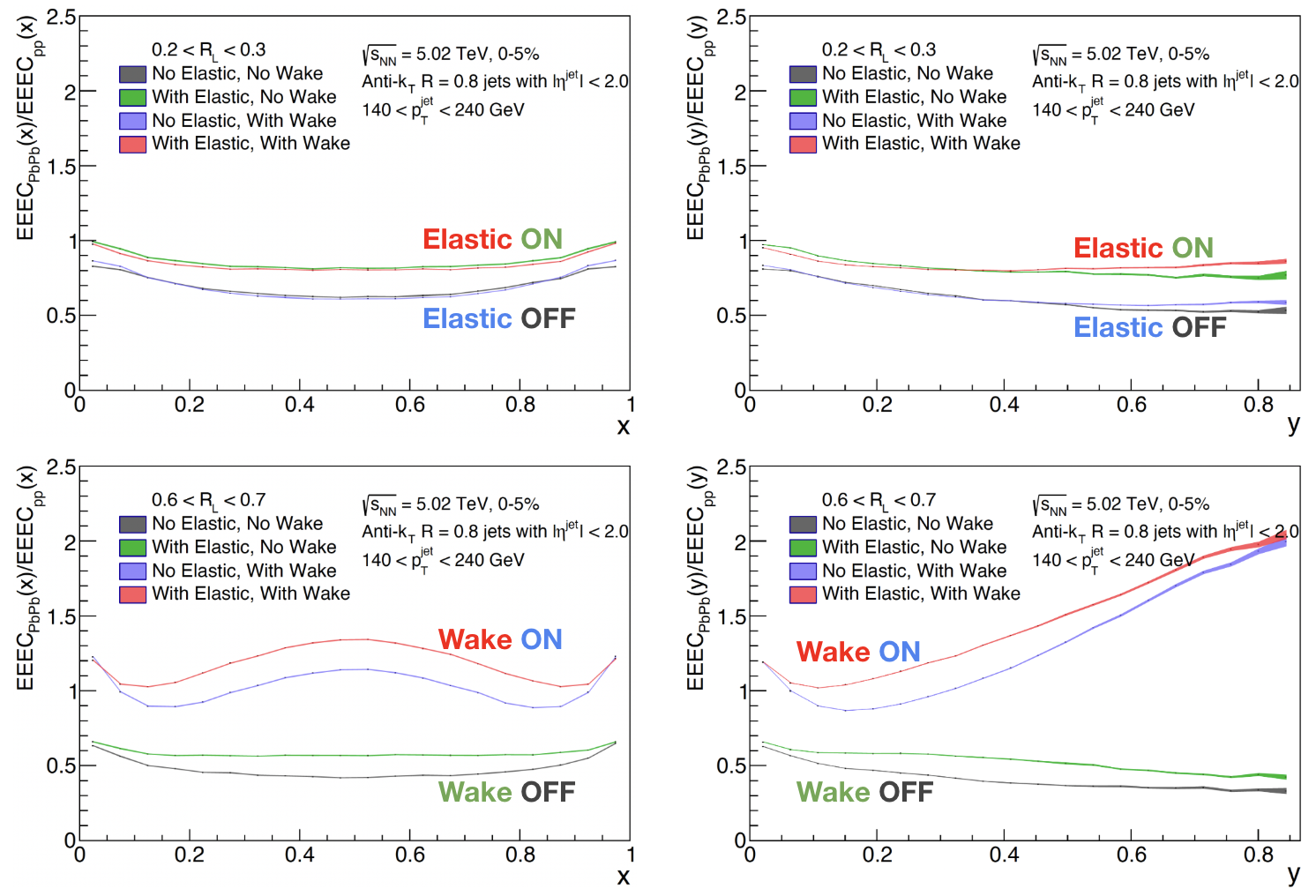}
\caption{PbPb/pp ratio of EEECs of anti-$k_t$ $R = 0.8$ jets with $140 < p_{\rm T}< 240$ GeV, projected onto the $x$-coordinate (left) and the $y$-coordinate (right) for $0.2 < R_L < 0.3$ (top) and $0.6 < R_L < 0.7$ (bottom). Hybrid Model curves are shown with/without elastic scattering and hadrons from the wakes of jets.}
\label{fig:e3c}       
\end{figure*}

Following Ref.~\cite{Bossi:2024qho}, we identify each particle-triplet that contributes to the EEEC as triangle with side lengths $R_S \le R_M \le R_L$. After rotating the configuration so that the side of length $R_L$ defines the horizontal base of the triangle and rescaling all distances by $R_L$, the third vertex of the triangle is located at an angular coordinate $(x,y)$. The other two vertices which define the longest side of the triangle are located at $(0,0)$ and $(1,0)$. The $(x,y)$ plane at fixed $R_L$ therefore parametrizes triangle \emph{shapes}: equilateral triangles cluster near $(x,y)\simeq(1/2,\sqrt{3}/2)$, flattened configurations lie near $y\simeq 0$, and squeezed topologies occur near $x\simeq 0$ or $1$. In this representation, the shape-dependent EEEC is a function of three angles --- $R_L$, $x$, and $y$ --- and cleanly separates an overall angular scale ($R_L$) from shape.

Fig.~\ref{fig:e3c} shows the PbPb/pp ratio of EEECs of anti-$k_t$ $R = 0.8$ jets with $140 < p_{\rm T}< 240$ GeV, projected onto the $x$-coordinate (left) and the $y$-coordinate (right) for two different ranges of $R_L$: $0.2 < R_L < 0.3$ (top) and $0.6 < R_L < 0.7$ (bottom). Two robust qualitative signatures emerge. First, consistent with findings in Ref.~\cite{Bossi:2024qho}, soft hadrons from the wake populate regions of phase space that are sparsely filled in vacuum, leading to an enhancement of \emph{equilateral} configurations at large $R_L$ (e.g. $0.6<R_L<0.7$). This appears as a steep rise around $(x,y)\sim(1/2,\sqrt{3}/2)$ in the bottom panels of Fig.~\ref{fig:e3c}. Second, elastic high momentum exchanges between jet-partons and quasi-particles in the medium result in angular deflection of energetic jet-partons. The imprints of elastic scatterings are most visible in the top panels of Fig.~\ref{fig:e3c}, at smaller $R_L$ where the wake is subdominant (e.g.\ $0.2<R_L<0.3$), producing a broad enhancement across $(x,y)$ rather than a purely equilateral rise.

The three-point EEEC thus furnishes a two-stage microscope: $R_L$ selects the angular \emph{scale} of the dynamics while $(x,y)$ isolates which triangular topologies, i.e. \textit{shapes} of the dynamics, are enhanced. The large equilateral enhancement at large $R_L$ diagnoses the angular structure of jet-induced wakes, while the more uniform uplift at smaller $R_L$ identifies elastic scatterings. Therefore, shape-dependent three-point EEECs provide access to high-granularity information on jet–medium interactions and the microscopic structure of QGP.
\newline

This research was supported in part by the U.S.~Department of Energy, Office of Science, Office of Nuclear Physics under grant contract numbers DE-SC0004168, DE-SC0011088 and DE-SC0011090.
ASK is supported by a National Science Foundation Graduate Research Fellowship Program under Grant No. 2141064.
IM is supported by a Sloan Research Fellowship.
DP is funded by the European Union's Horizon 2020 research and innovation program under the Marie Sk\l odowska-Curie grant agreement No 101155036 (AntScat), by the European Research Council project ERC-2018-ADG-835105 YoctoLHC, by the Spanish Research State Agency under project 
PID2020-119632GB-I00, by Xunta de Galicia (CIGUS Network of Research Centres) and the European Union, and by Unidad de Excelencia Mar\'ia de Maetzu under project CEX2023-001318-M.
DP also acknowledges support from the the Ram\'on y
Cajal fellowship RYC2023-044989-I.
KR acknowledges the hospitality of the CERN Theory Department and the Aspen Center for Physics, which is supported by National Science Foundation grant PHY-2210452.

\bibliography{bibliography.bib} 
%
%

\end{document}